\documentclass{emulateapj}

\slugcomment{Accepted for publication in The Astrophysical Journal}

\shorttitle{Space density of X-ray-selected AGN}
\shortauthors{Silverman et al.}

\begin{document}


\title{Co-moving space density of X-ray-selected Active Galactic Nuclei}


\author{J. D. Silverman\altaffilmark{1,2,7}, P.
J. Green\altaffilmark{1}, W. A. Barkhouse\altaffilmark{1},
R. A. Cameron\altaffilmark{1}, C. Foltz\altaffilmark{5}, B.
T. Jannuzi\altaffilmark{4}, D.-W. Kim\altaffilmark{1}, M.
Kim\altaffilmark{1}, A. Mossman\altaffilmark{1}, H.
Tananbaum\altaffilmark{1}, B. J. Wilkes\altaffilmark{1}, M.
G. Smith\altaffilmark{6}, R.C. Smith\altaffilmark{6} and P.
S. Smith\altaffilmark{3}}

\altaffiltext{1}{Harvard-Smithsonian Center for Astrophysics, 60 Garden Street, Cambridge, MA 02138}

\altaffiltext{2}{Astronomy Department, University of Virginia,
P.O. Box 3818, Charlottesville, VA, 22903-0818}

\altaffiltext{3}{Steward Observatory, The University of Arizona, Tucson, AZ 85721}

\altaffiltext{4}{National Optical Astronomy Observatory, P.O. Box 26732, Tucson, AZ, 85726-6732}

\altaffiltext{5}{National Science Foundation, 4201 Wilson Blvd., Arlington, VA, 22230}

\altaffiltext{6}{Cerro Tololo Inter-American Observatory, National Optical Astronomical Observatory, Casilla 603, La Serena, Chile}

\altaffiltext{7}{jsilverman@mpe.mpg.de}


\begin{abstract}

For measurement of the AGN luminosity function and its evolution,
X-ray selection samples all types of AGN and provides reduced
obscuration bias in comparison with UV-excess or optical surveys.  The
apparent decline in optically-selected quasars above $z\sim3$ may be
strongly affected by such a bias.  The {\em Chandra} Multiwavelength
Project (ChaMP) is characterizing serendipitously detected X-ray
sources in a large number of fields with archival {\em Chandra}
imaging. We present a preliminary measure of the co-moving space
density using a sample of 311 AGN found in 23 ChaMP fields ($\sim1.8$
deg$^{2}$) supplemented with 57 X-ray bright AGN from the CDF-N and
CDF-S. Within our X-ray flux ($f_{\rm 0.3-8.0~keV}>4\times10^{-15}$
erg cm$^{-2}$ s$^{-1}$) and optical magnitude ($r^{\prime}<22.5$)
limits, our sample includes 14 broad emission line AGN at $z>3$.
Using this X-ray selected sample, we detect a turnover in the
co-moving space density of luminous type 1 AGN (log $L_{\rm X}>44.5$;
units erg s$^{-1}$; measured in the 0.3--8.0 keV band and corrected
for Galactic absorption) at $z>2.5$.  Our X-ray sample is the first to
show a behavior similar to the well established evolution of the
optical quasar luminosity function.  A larger sample of high redshift
AGN and with a greater fraction of identified sources, either
spectroscopic or photometric, at faint optical magnitudes
($r^{\prime}>22.5$) are required to remove the remaining uncertainty
in our measure of the X-ray luminosity function, particularly given
the possibility that AGNs might be more easily obscured optically at
high redshift.  We confirm that for $z<1$, lower luminosity AGN
(log~$L_{\rm X}<44.5$) are more prevalent by more than an order of
magnitude than those with high luminosity.  We have combined the {\em
Chandra} sample with AGN from the {\em ROSAT} surveys to present a
measure of the space density of luminous type 1 AGN in the soft X-ray
band (0.5--2.0 keV) which confirms the broad band turnover described
above.

\end{abstract}




\section{Introduction}

Optical surveys have measured the evolution of QSOs out to $z\sim6$
\citep{fa04}.  The most dramatic feature found is the rise and fall of
the co-moving space density with peak activity at $z\sim2.5$.  A
systematic decrease in luminosity from $z\sim2$ to the present is
evident with very few intrinsically bright QSOs in the local universe
(e.g., \citealt{cr04}).  This fading of the QSO population is
attributed to a decreased fuel supply and/or fueling rate (e.g.,
\citealt{ca00, ka00}).  The dropoff in the space density at $z>3$ 
(e.g., \citealt{wa94,sc95,fa01,wo03}) could
represent the growth phase of supermassive black holes (SMBHs),
possibly under a veil of obscuration \citep{fa99}.

It has become clear in the past decade that X-ray selection of AGN
offers many benefits over detection methods in other wavebands.
X-rays can penetrate large absorbing columns of gas which can
effectively hide an accreting black hole in the optical.  The
existence of a significant, missed population of obscured AGN is
required to explain the spectral shape of the Cosmic X-ray Background
(CXRB; e.g., \citealt{gi01}).  The current generation of X-ray
observatories ({\em Chandra}, {\em XMM-Newton}) are probing the faint
source population responsible for the bulk of the CXRB (e.g.,
\citealt{al03,ro02,ha01}).

X-ray surveys with {\em Chandra}, {\em XMM-Newton} and $ASCA$ are
refining our knowledge of the AGN X-ray luminosity function.  The
peak, in the co-moving space density of low luminosity
($L_{2.0-8.0~{\rm keV}}<10^{44}$ erg s$^{-1}$) AGN found in hard
(2--10 keV) X-ray surveys (\citealt{ba03,co03,fi03,st03,ue03}), occurs
at $z\sim1$.  This contrasts starkly with the behavior of the luminous
($L_{2.0-8.0~{\rm keV}}>10^{44}$ erg s$^{-1}$) QSOs which are most
prevalent at $z\sim2.5$.  Using a highly complete sample of 941 AGN
selected in the soft band, \citet{ha05} show the
same luminosity dependence and extend the space density measurement of
low luminosity AGN ($L_{X}<10^{44}$ erg s$^{-1}$) out to $z\sim3$.
These latest results clearly require a luminosity dependent density
evolution model \citep{mi00} in contrast to a 'pure' luminosity
evolution model, descriptive of optically selected QSOs \citep{cr04}.
With these results, semi-analytic models \citep{me04} posit
that the more luminous QSOs illuminate early epochs when most of the
massive galaxies are forming ($z>2$), thereby inducing high accretion
rates, whereas lower luminosity AGN dominate at a later period ($z<2$)
when most of the galaxies have fully assembled.

X-ray selected QSOs from {\em ROSAT} have hinted at a constant space
density between $1.5<z<4.5$ \citep{mi00} for the most luminous AGN
(log $L_{\rm 0.3-8.0~keV} > 44.5$).  The lack of a decline in the
space density at $z>3$, in contrast to the behavior seen in optical
surveys, could be evidence for obscuration at early epochs.  We can
test this model (Fabian 1999) by measuring the luminosity function and
co-moving space density of X-ray-selected AGN over a wide area to
compile a significant sample at $z>3$.  Unfortunately, the {\em ROSAT}
sample includes only 7 QSOs at these high redshifts.  Of these, only 5
QSOs have luminosities high enough to be detected over a broad range
of redshift ($0<z<5$) and thereby included in the measurement of the
co-moving space density \citep{mi00}.  While the {\em Chandra} and
{\em XMM-Newton} Deep field observations have great scientific merit,
their narrow area provides only 4 high luminosity (log $L_{\rm
0.3-8.0~keV} > 44.5$) AGN at $z>3$.  Recent measurement of the hard
X-ray luminosity function (\citealt{ba03,fi03,st03,ue03}) have limited
numbers of AGN at these high redshifts.  A wider area survey is needed
to compile a significant sample of highly luminous QSOs with redshifts
greater than 3.

We present a preliminary measure of the co-moving space density of
X-ray selected AGN out to $z\sim4$ from the {\em Chandra}
Multiwavelength Project (ChaMP).  We supplement the ChaMP sample with
AGN of comparable X-ray and optical fluxes from the {\em Chandra} Deep
Field North (CDF-N) and South (CDF-S) to boost our spectroscopic
completeness at faint optical magnitudes and include one additional
$z>3$ QSO.  We merge the subsample of these AGN detected in the soft
(0.5--2.0 keV) band with those from {\em ROSAT} to directly compare
with Miyaji et al. (2000, 2001).  A full presentation of the ChaMP
X-ray luminosity function is forthcoming (Silverman et al. 2005, in
preparation).  We assume H$_{\circ}=70$ km s$^{-1}$ Mpc$^{-1}$,
$\Omega_{\Lambda}=0.7$, and $\Omega_{\rm{M}}=0.3$ with the exception
of Section~\ref{sxlf}.

\section{Chandra Multiwavelength Project}

The ChaMP (\citealt{ki04a,gr04}) is carrying out an extragalactic
X-ray survey encompassing 10 deg$^{2}$ using serendipitous detections
in archival {\em Chandra} fields.  $Chandra's$ small point spread
function ($\sim1^{\prime\prime}$ resolution on-axis) and low
background allow sources to be detected to fainter flux levels
($\sim10^{-15}$ erg cm$^{-2}$ s$^{-1}$) than any X-ray observatory
past or present, thus enabling the detection of high redshift
($z\sim5$) AGN (\citealt{tr04,ba03,ca03,si02}).  For the present
study, we use the full {\em Chandra} energy range (0.3--8.0 keV) to
detect the absorbed sources missed by previous optical, UV or soft
X-ray surveys and take advantage of the high collecting area at soft
energies to detect the faint, high redshift AGN.  At $z>3$, we are
sensitive to AGN with absorbing columns up to $\sim10^{23}$ cm$^{-2}$
due to the favorable $k$-correction.  With a sample of AGN selected
from a large number of non-contiguous {\em Chandra} fields that reach
similar depths ($\sim10^{-15}$ erg cm$^{-2}$ s$^{-1}$), we effectively
smooth out any effects from large scale structures such as those
evident in the CDF-S \citep{gi03}.

We have chosen 23 {\em Chandra} fields (1.8 deg$^{2}$) for which we have
acquired extensive followup optical imaging and spectroscopy.  The
deepest observations have exposure times that are sensitive to sources
with $f_{0.3-8.0\rm{keV}}>8\times10^{-16}$ erg cm$^{-2}$ s$^{-1}$.  A
full description of the ChaMP image reduction and analysis pipeline
XPIPE can be found in \citet{ki04b}.  With our 4m MOSAIC optical
imaging, we are able to identify counterparts to the {\em Chandra} sources
down to $r^{\prime}\sim25$ \citep{gr04}.  We acquire optical
imaging in three ($g^{\prime}, r^{\prime},$ and $i^{\prime}$) Sloan
Digital Sky Survey (SDSS) filters \citep{fu96}.  Optical
colors provide preliminary source classification and crude photometric
redshifts.  These diagnostics are required to characterize the
optically faint X-ray sources that cannot be identified with
spectroscopy and in particular, their influence on the X-ray
luminosity function.  The use of the SDSS photometric system allows
more direct comparison between the ChaMP and the SDSS AGN surveys
\citep{si04}.  Optical spectroscopic followup currently
focuses on identifying counterparts with $r^{\prime}<22.5$ for which
spectra can be acquired on a 4-6m (i.e. MMT, Magellan, WIYN, CTIO
Blanco) class telescope.  To date, we have spectroscopically
classified a sample of 358 AGN detected in the broad band (0.3--8.0
keV) in these 23 ChaMP fields.

\section{X-ray sensitivity and area coverage}
\label{area}

There are complications with measuring the X-ray luminosity function
that must be put into context.  The difficulty in focusing X-rays
results in a point source function (PSF) and flux sensitivity that
varies across the field of view.  The PSF degrades as a function of
off-axis angle, decreasing the flux sensitivity.  In terms of
measuring the luminosity function, it is not a trivial task to
determine the incompleteness at the faintest X-ray fluxes and the
actual sky area over which a source of a specific flux would be
detected.  In addition, the exposure times of these Chandra fields
range from 17 to 114 ksec generating a wide range of limiting fluxes
even on-axis.

\begin{figure}
\epsscale{1.0}
\plotone{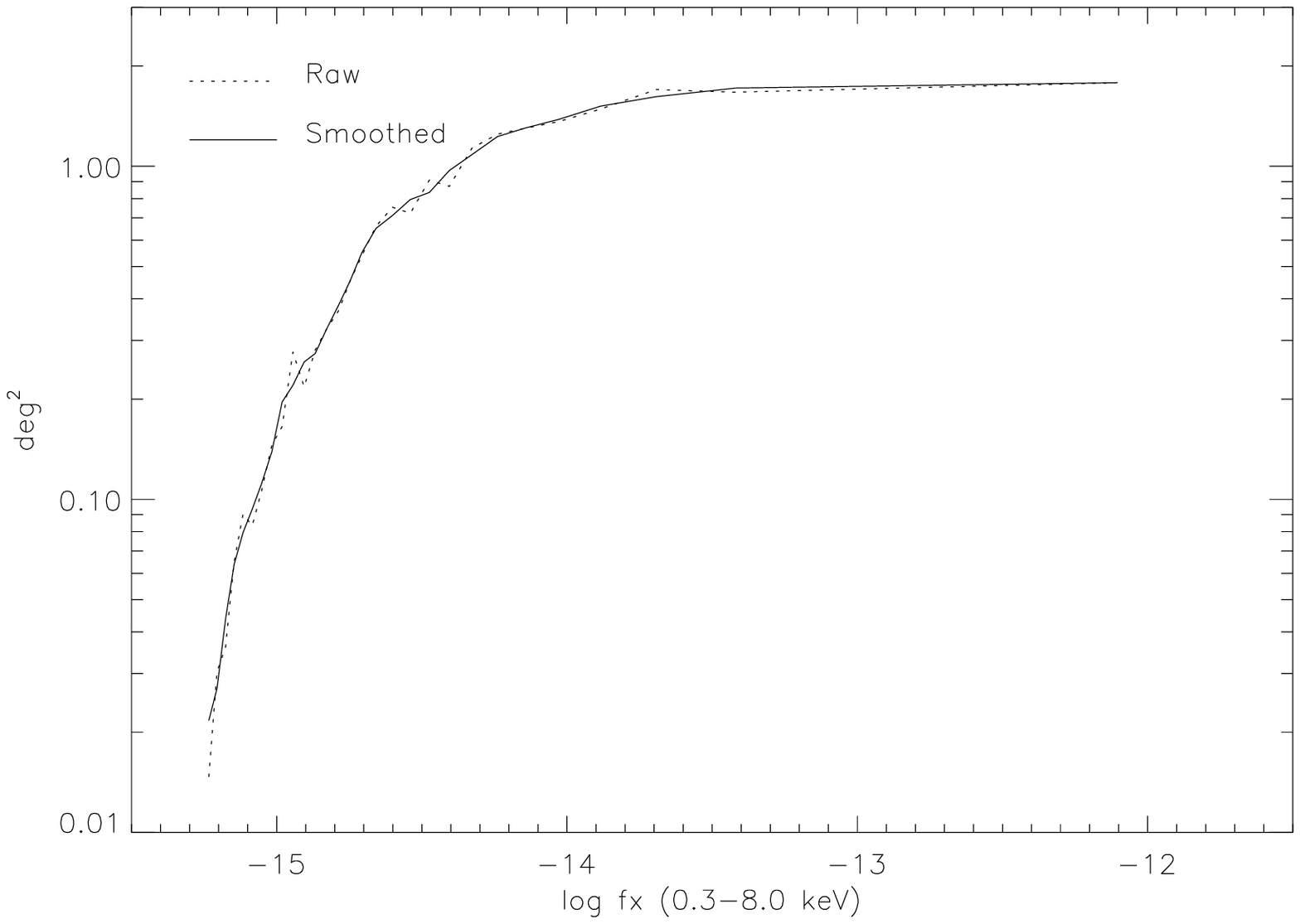}
\epsscale{1.10}
\plotone{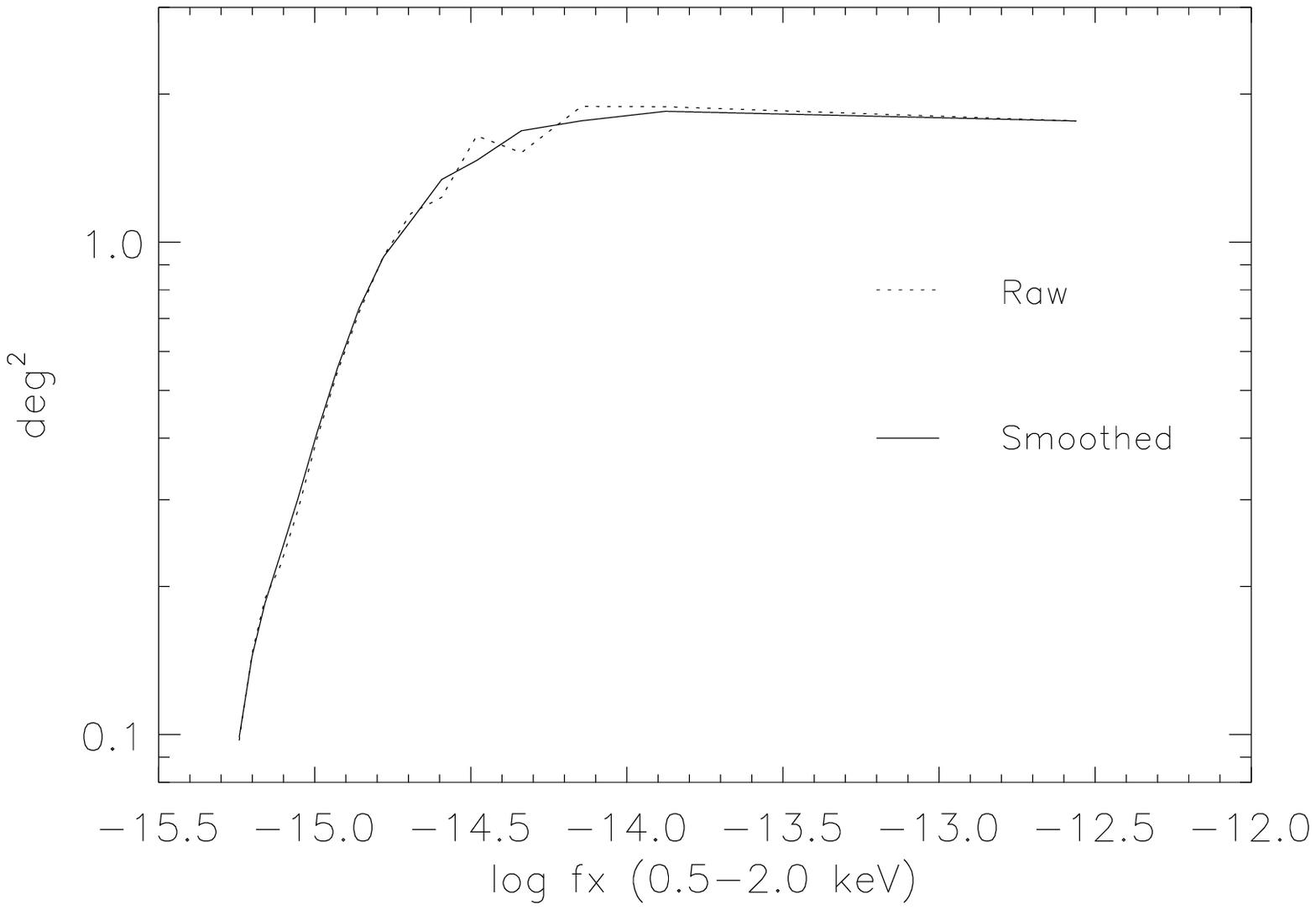}
\caption{Area coverage as a function of broad (0.3--8.0 keV; $top$)
and soft (0.5--2.0 keV; $bottom$) band X-ray flux for 23 ChaMP fields.
A smoothed curve is generated since small scale variations or sharp
features are a result of the limited number of simulated sources and
not real.}
\label{fig:area}
\end{figure}

To characterize the sensitivity, completeness, and sky area coverage
as a function of X-ray flux, a series of simulations were performed.
The full details will be presented in an upcoming ChaMP X-ray analysis
paper (M. Kim, D.-W Kim et al., in preparation).  The simulations
consists of three parts, 1) generating artificial X-ray sources with
MARX (MARX Technical Manual\footnote{http://space.mit.edu/CXC/MARX})
and adding them to real X-ray images, 2) detecting these artificial
sources by {\tt wavdetect} and extracting source properties with XPIPE
identically as performed for actual sources, and 3) estimating the sky
area coverage by comparing the input and detected source properties as
a function of flux.  The simulations are restricted to specific CCDs
for ACIS-I (I0, I1, I2, and I3) and ACIS-S (I2, I3, S2, and S3).
These CCDs are closest to the aimpoint for each observation.  Sources far
off-axis ($\Theta>12\arcmin$) are excluded since the flux sensitivity
is low and the PSF is degraded.  These simulations allow determination
of corrections for the source detection incompleteness at faint flux
levels quantified in the first ChaMP X-ray analysis paper
\citep{ki04b}.  In Figure~\ref{fig:area}, we show the sky coverage
determined from the simulations using the broad (0.3--8.0 keV) and
soft (0.5--2.0 keV) band source detections for 23 ChaMP fields.  We
are surveying a sky area of 1.8 deg$^{2}$ for the brightest sources.
The sky area falls below 0.1 deg$^{2}$ at the faintest flux levels.

\section{AGN selection}

\label{selection}

\begin{figure}
\epsscale{1.1}
\plotone{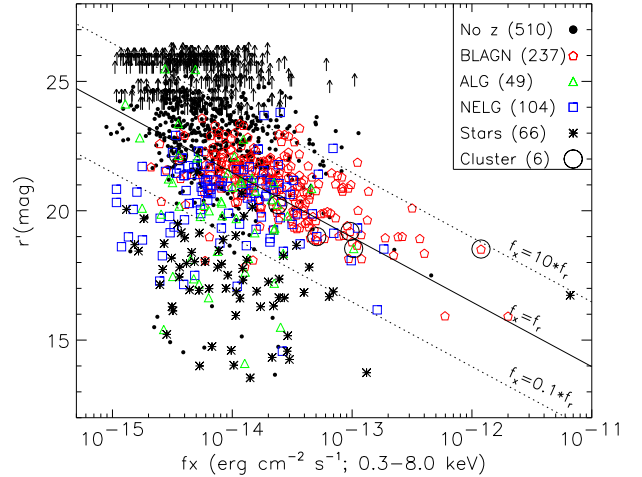}
\caption{Optical magnitude ($r^{\prime}$) vs. X-ray flux (0.3--8
 keV) of X-ray sources found in 23 ChaMP fields.  Optical
 spectroscopic classifications are indicated with the sample size.
 X-ray sources with no optical counterparts are shown by an arrow
 placed at the magnitude for a 5$\sigma$ detection from our optical
 imaging.}
\label{xo:fig}
\end{figure}

In these ChaMP fields, we find a diversity of objects (AGN, clusters,
galaxies, and stars), although 85\% of them are attributed to an AGN
(\citealt{gr04,si04}).  We show the optical magnitude ($r^{\prime}$)
as a function of 0.3--8.0 keV X-ray flux for the sources detected in
23 ChaMP fields (Figure~\ref{xo:fig}).  We only include sources with
greater than 9.5 net counts.  Although sources with counts as low as 2
may be significant detections as a result of {\em Chandra}'s low
background, we use a higher count limit to ensure that we have
well-measured X-ray fluxes within our sample.  In Figure~\ref{xo:fig}
we label sources as classified from our optical spectroscopy.
Extragalactic objects with strong emission lines (W$_{\lambda}>5$
\rm{${\rm \AA}$}) are labelled as either Broad Line AGN (BLAGN; FWHM
$>$ 1000 km s$^{-1}$) or Narrow Emission Line Galaxy (NELG; FWHM $<$
1000 km s$^{-1}$).  The BLAGN are equivalent to type 1 AGN.  We
loosely refer to the NELG with $L_{\rm X}>10^{42}$ erg s$^{-1}$ as
type 2 AGN since optical spectra may not cover all the emission
lines needed to confirm photoionization from a non-thermal source.
Extragalactic counter-parts with weak emission line (W$_{\lambda}<5$
\rm{${\rm \AA}$}) or pure absorption line spectra are classified as 
Absorption Line Galaxy (ALG).  In addition, a handful of stars are 
identified at faint X-ray fluxes.  Six clusters have been found based 
primarily on their extended X-ray emission.

We calculate the rest frame (0.3--8.0 keV; erg s$^{-1}$) X-ray
luminosity for each extragalactic source, after correction for
Galactic absorption.  The conversion from X-ray count rate to flux
units (erg cm$^{-2}$ s$^{-1}$ ) is determined from simulated
detections on each CCD of a source with a powerlaw
spectrum\footnote{The ChaMP XPIPE \citep{ki04b} provides energy
conversion factors (ECF) for two models with $\Gamma=1.7$ and
$\Gamma=1.4$.  We chose the former since the photon index more closely
resembles the majority of the X-ray source detections with
spectroscopic identification.} ($f_{E}\propto
E^{-(\Gamma-1)}$;$\Gamma=1.7$) and galactic absorption \citep{di90}.
The observed luminosity is converted to the rest frame assuming a
powerlaw spectrum with the photon index ($\Gamma$) set to the average
value for that object type (BLAGN, NELG, ALG) based upon our X-ray
spectral fit results (Aldcroft et al.\ in preparation). We find that
the BLAGN have a mean spectral index ($<\Gamma>$) of 1.9 while the
NELG have $<\Gamma>$=0.9 and ALG have flatter slopes $<\Gamma>$=1.5.
We do not apply a correction for intrinsic absorption due to the
uncertainty in the absorbing columns for most of these AGN as a result
of their low X-ray counts.

\begin{figure}
\epsscale{1.1}
\plotone{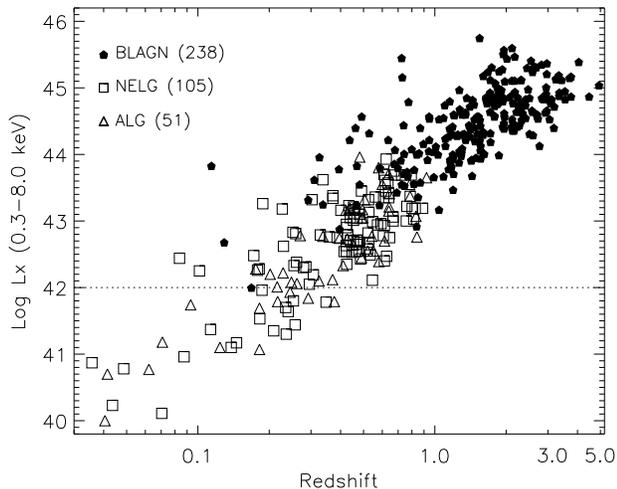}
\caption{X-ray luminosity vs. redshift of 394 extragalactic objects
  found in 23 ChaMP fields (with the exception of 6 clusters).  The
  dashed line marks the minimum luminosity required for the AGN
  sample.  Optical spectroscopic classifications are shown as either
  BLAGN, NELG or ALG.} 
\label{champ_lz}
\end{figure}

To construct a pure AGN sample, we require the derived (rest frame
0.3--8.0~keV) luminosity to exceed 10$^{42}$ erg s$^{-1}$ thereby
excluding any sources that contain a significant stellar or hot ISM
component.  As shown in Figure~\ref{champ_lz}, we see that 92\% of the
extragalactic sources in the ChaMP not associated with an extended
X-ray emitting cluster satisfy this criterion.  We currently have not
identified any NELG and ALG above a redshift of 1 in the ChaMP.  Our
optical spectroscopic limit ($r^{\prime}\sim22$) precludes
classification of objects with heavy optical extinction beyond this
redshift, since a $\sim5L_{\star}$ elliptical host galaxy has
$r^{\prime}\sim23$ at $z\sim1$.  We detect many optical type 2 AGN
with $L_{\rm X}<10^{44}$ erg s$^{-1}$ but no luminous ($L_{\rm
X}>10^{44}$ erg s$^{-1}$) optical type 2 QSOs.  We do find 6\% of the
BLAGN to be X-ray absorbed with a substantial fraction having reddened
optical colors which may prevent them from being selected in optical
surveys \citep{si04}.  These red BLAGN may be similar to those found
by the Two Micron All Sky Survey \citep{fr04} that tend to have
significant X-ray absorbing columns of gas \citep{wi02}.  Deep optical
followup of a broadband X-ray-selected sample will help determine the
fraction of such red quasars to the space density of X-ray luminous
AGN.

\section{Survey completeness}
\label{complete}

\begin{figure}
\epsscale{1.1}
\plotone{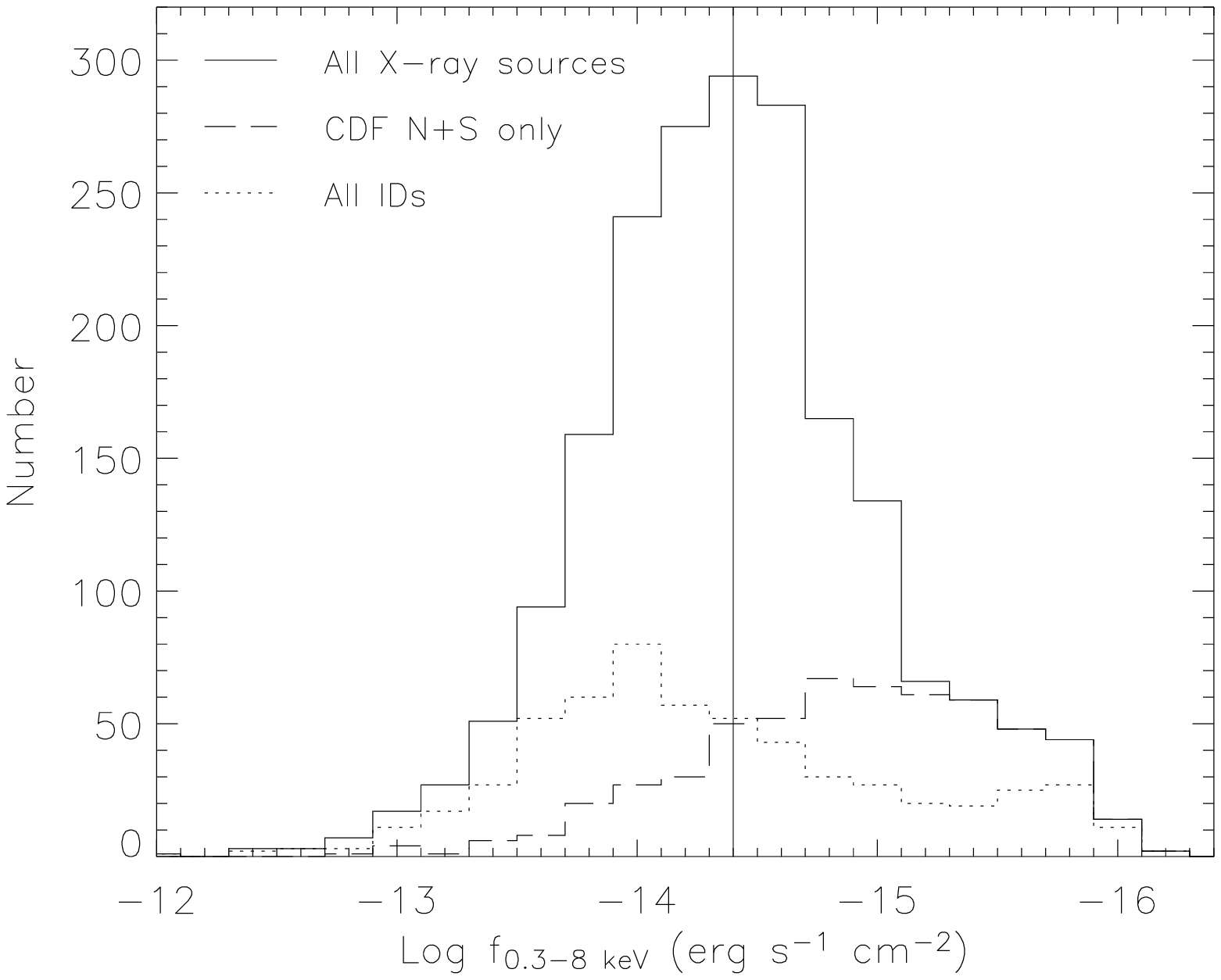}
\plotone{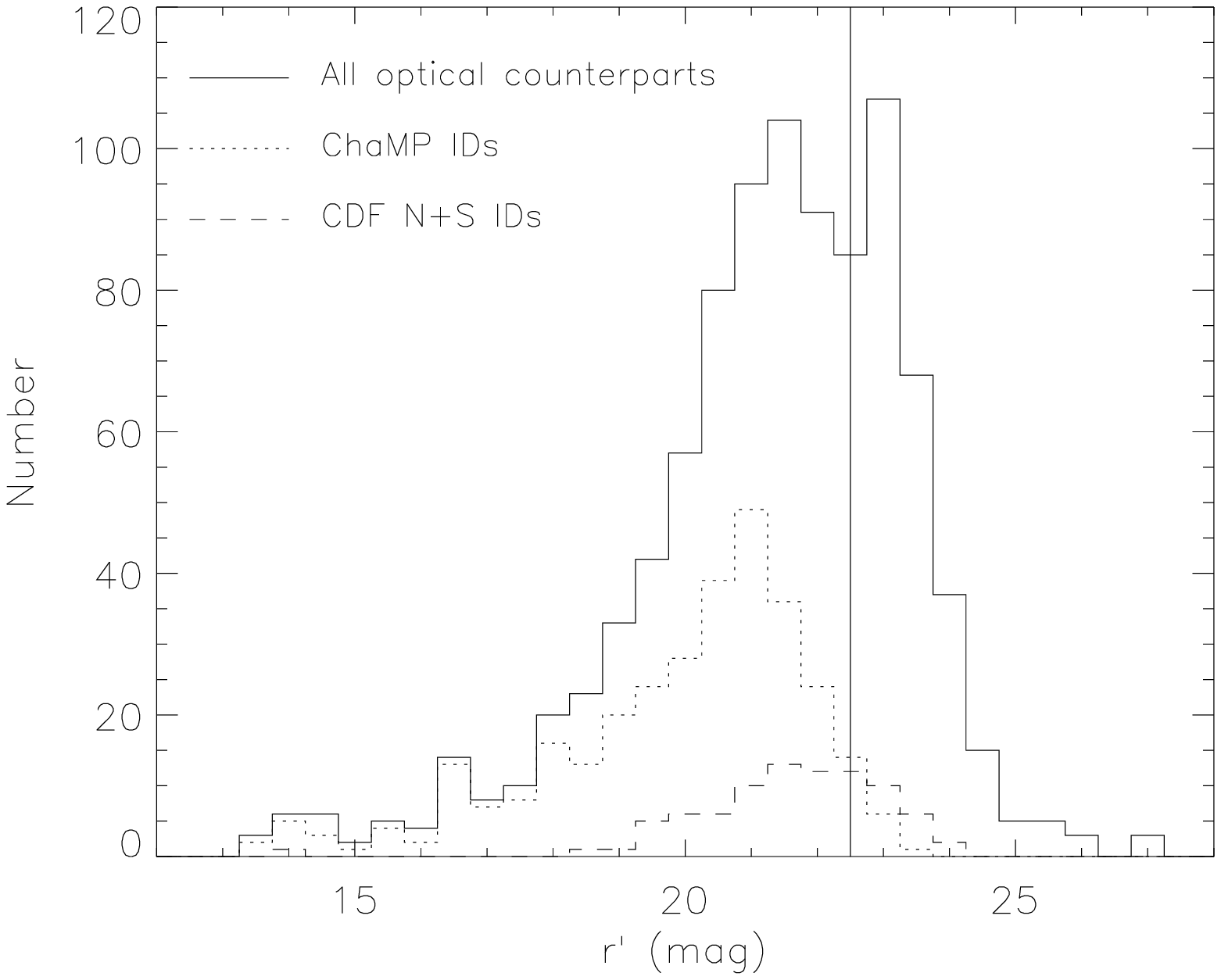}
\caption{$top$ X-ray (0.3-8.0 keV) flux distribution of 1989 sources
in the ChaMP and CDF N+S surveys (solid histogram).  Those sources
from only the CDF N+S are shown by the long dashed line.  Those with
optical spectroscopic identifications are shown by the short dashed
line.  The vertical line marks the X-ray flux limit of $f_{\rm
0.3-8.0~keV}>4\times10^{-15}$ erg cm$^{-2}$ s$^{-1}$.  $bottom$
Optical magnitude ($r^{\prime}$) distribution of counterparts to X-ray
sources with fluxes above our limit.  Counterparts with spectroscopic
identification are shown with a short (ChaMP) or long (CDF N+S) dashed
line.  The vertical line marks our chosen optical magnitude limit.}
\label{xodistr}
\end{figure}

We have supplemented our ChaMP sample with X-ray sources from
the CDF-N \citep{br01,ba02} and CDF-S
\citep{sz04} with X-ray fluxes above our chosen limit
(see below; $f_{\rm 0.3-8.0~keV}>4\times10^{-15}$ erg cm$^{-2}$
s$^{-1}$).  These samples allow us to boost the number of
spectroscopically identified sources at faint optical magnitudes
($r^{\prime}>21$) and include any additional $z>3$ AGN.  These are
currently the only surveys with published optical spectroscopic
identifications at moderate completeness levels for sources with flux
levels comparable to the ChaMP.  In Figure~\ref{xodistr} $top$, we
plot the flux distribution of X-ray sources found in these 25
{\em Chandra} fields.  The CDF-N and CDF-S clearly contribute most of the
faint sources (log $f_X<-15$).  In Figure~\ref{xodistr} $bottom$, we
show the optical magnitude distribution of counterparts to sources
with X-ray fluxes above our chosen limit and those with optical
spectroscopic identifications.  Optical magnitudes for the CDF N+S
sources were converted from the Johnson (B, R) to the SDSS photometric
system using the transformation in \citet{fu96}.  By including these
two deep Chandra fields, our area coverage is larger than that shown
in Figure~\ref{fig:area}.  We add a fixed area of 0.22 deg$^2$ since
the area coverage in these two fields does not change above this
bright flux limit by more than 3\%.

\begin{figure}
\epsscale{1.2}
\plotone{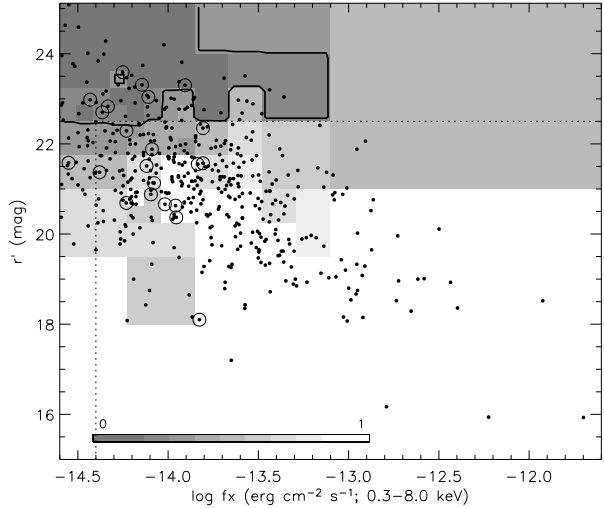}
\caption{Adaptively binned spectroscopic completeness map.
 The shading scale ranges from zero (lightest) to 100\% (darkest)
 spectroscopically classified.
 Only spectroscopically identified AGN are shown with a black dot,
 those at $z>3$ circled.  Dashed lines show our chosen flux limits.  A
 solid contour line has been overplotted to mark the 30\% completeness
 level.}
\label{xo:adapt}
\end{figure}

To evaluate the status of our optical followup and determine effective
survey limits, we measure the fraction of X-ray sources identified
with optical spectroscopic observations as a function of X-ray flux
and optical magnitude.  Since the X-ray sources do not fully sample
the $f_{\rm{X}}-r^{\prime}$ plane (Figure~\ref{xo:fig}), we have
implemented an adaptive binning scheme identical to that of
\citet{sa01} to generate a completeness map as a function of X-ray and
optical flux for the combined sample (ChaMP + CDF-N + CDF-S).  We
create an array with 64$\times$64 elements covering the
$f_{\rm{X}}-r^{\prime}$ plane (Figure~\ref{xo:fig}).  The identified
fraction $f_{\rm ID}$ is measured at each element as the number of
identified objects divided by the number of X-ray sources.  If 10 or
more X-ray sources are in a single element, then we assign $f_{\rm
ID}$ as the completeness level.  We then increase the bin size by a
factor of two.  If the binned element has 10 or more objects, the
individual elements not previously assigned an identified fraction are
set to this value.  This procedure is iterated until the final
binned element equals the size of the full array.  In
Figure~\ref{xo:adapt}, we plot the final array (i.e. completeness map)
as a grey scale image.

Based on Figure~\ref{xo:adapt}, we set the flux limits of our AGN
sample.  We choose a single X-ray flux limit at $f_{\rm
0.3-8.0~keV}>4\times10^{-15}$ erg cm$^{-2}$ s$^{-1}$ that is brighter
than the faintest X-ray detections in the ChaMP.  This limit ensures
that we minimize the bias towards optically bright AGN at the faintest
X-ray flux limits.  In order to limit our completeness correction to
$<3.3$, we set an optical magnitude limit at a level where at least
30\% of the sources are identified, this leads to a magnitude limit of
22.5.  With the remaining sample, the incompleteness level is
used to correct the measurement of the space density (Section~\ref{xlf}).

\begin{figure}
\epsscale{1.1}
\plotone{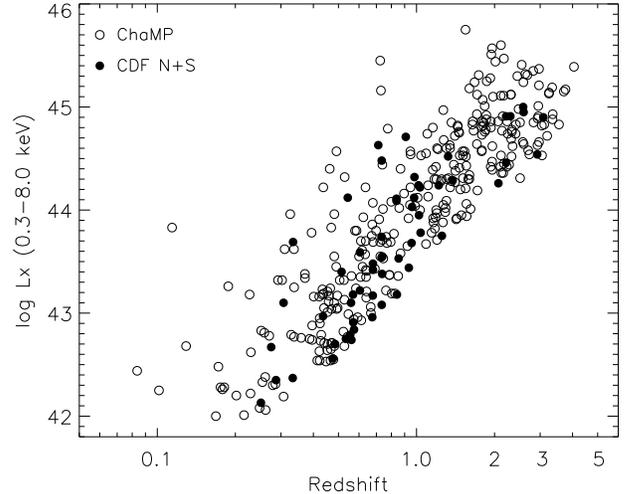}
\caption{X-ray luminosity, redshift distribution of 368 AGN (311 ChaMP + 57 CDF N+S) with
fluxes above our chosen limits.}
\label{lzdistr}
\end{figure}

In Figure~\ref{xo:adapt}, we plot the location of AGN, including those
from the {\em Chandra} Deep Fields in the $f_{\rm{X}}-r^{\prime}$ plane.
Since a significant number of AGN fall in regions of low completeness
at faint fluxes, we must exclude
 8 of the 22 AGN with $z>3$ and log $f_X>-14.6$ (circled points).
In particular, the highest redshift AGN (CXOMP J213945.0-234655;
$z=4.93$; \citealt{si02}) found by the ChaMP has been excluded
not because of its X-ray flux ($f_{\rm
0.3-8.0~keV}=4.32\times10^{-15}$ erg cm$^{-2}$ s$^{-1}$) but rather
due to its optical magnitude ($r^{\prime}=22.7$) which falls below our
limit.  We do have a significant sample of 368 AGN above our flux
limits with 14 (13 from ChaMP) at $z>3$, almost a factor of
three larger than the high redshift sample available from {\em ROSAT}.
In Figure~\ref{lzdistr}, we plot the $L_{X}-z$ distribution of the
total sample of 368 AGN selected in the 0.3-8.0 keV band.

\section{Co-moving space density}
\label{xlf}

We implement the 1/V$_{a}$ method \citep{sc68} to generate an estimate
to the co-moving space density ($n$ in units of Mpc$^{-3}$;
Equation~\ref{eq:va}) in fixed redshift and luminosity intervals.  For
each $L-z$ bin, Equation~\ref{eq:va} is summed over all AGN (N) from
our sample (Figure~\ref{lzdistr}) that fall within this bin.

\begin{equation}
n=\sum_{i=1}^{N} \frac{1}{f_{\rm ID}V_{a}(i)
}
\label{eq:va}
\end{equation}

\noindent As mentioned in Section~\ref{area}, our survey
has non-contiguous sky coverage with varying flux limits.  To account
for this, we determine the accessible volume $V_{a}$
(Equation~\ref{eq:vol}) over which an individual AGN $i$ will be
included in our sample, given the X-ray and optical
flux limits.

\begin{equation}
V_{\rm a} = \int_{z_1}^{z_2}\frac{dV_{c}}{dz}\,dzd\Omega
\label{eq:vol}
\end{equation}

\noindent We calculate the X-ray and optical limiting redshifts
($z_{lim}^{X}$, $z_{lim}^{O}$) for each object.  The integrand in
Equation~\ref{eq:vol} is summed from $z_1$ to $z_2$ where $z_2$ is the
smaller of $z_{lim}^{X}$ and $z_{lim}^{O}$ if the AGN cannot be
detected over the full redshift interval.  The solid angle (d$\Omega$)
is a function of X-ray flux as shown in Figure~\ref{fig:area} plus an
additional 0.22 deg$^2$ contributed by the Chandra Deep Fields
observations.  A correction factor ($f_{\rm ID}$), determined from the
location of an individual AGN in the $f_{\rm{X}}-r^{\prime}$ plane
(Figure~\ref{xo:adapt}), compensates for the incompleteness in our
spectroscopic identifications as detailed in the previous section.
Each AGN contributes ($f_{\rm ID}$$V_{a}$)$^{-1}$ to the space density
in a specific luminosity and redshift interval.  We estimate 1$\sigma$
errors based on a Poisson distribution due to the small number of
objects per redshift bin.  While there are known biases inherent in
the 1/V$_{a}$ method (e.g. \citealt{mi01,pa00}), the overall
evolutionary trends can be discerned (e.g. \citealt{ba03,co03,fi03}).
A detailed X-ray luminosity function, will be presented in a future
publication (Silverman et al.\ in preparation), while implementing
more sophisticated analysis techniques (e.g. maximum likelihood
method).

\subsection{Broad band (0.3--8.0 keV)}
\label{bxlf}

\begin{figure}
\epsscale{1.1}
\plotone{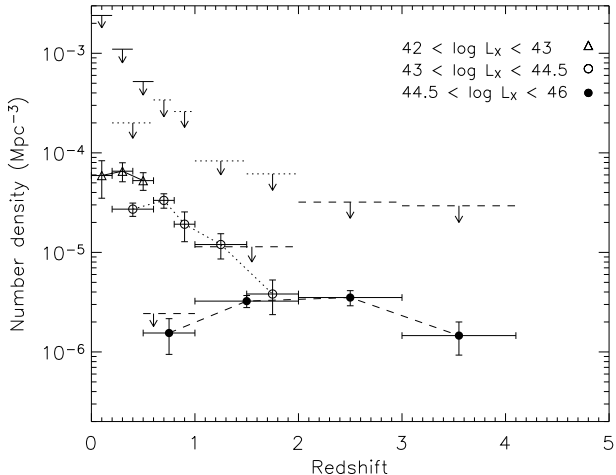}
\caption{Co-moving space density of AGN selected in the broad
(0.3--8.0 keV) band.  Vertical bars are estimates of the $1\sigma$
error.  Horizontal bars, centered on each data point, mark the
redshift bin size.  Horizontal bars with a downward arrow show the
highly conservative upper limits that take into account the optically
faint ($r^{\prime}>22.5$), unidentified X-ray sources.  The upper
limits correspond to the data which have the same line type connecting
the points.}
\label{bxlf:density}
\end{figure}

We measure the co-moving space density (Equation~\ref{eq:va}) of AGN
detected in the broad band from 25 {\em Chandra} fields.  In
Figure~\ref{bxlf:density}, the number density is plotted in three
luminosity bins ($42.0<{\rm log}~L_{\rm X}<43.0$; $43.0<{\rm
log}~L_{\rm X}<44.5$; $44.5<{\rm log}~L_{\rm X}<46.0$; $L$ is in units
of erg s$^{-1}$) as a function of redshift.  We must keep in mind that
we are limited to the detection of obscured AGN (type 2) at
luminosities below $10^{44}$ erg s$^{-1}$ and redshifts below 1 due to
our optical spectroscopic limit as shown in Figure~\ref{champ_lz}.  As
a result, the highest luminosity range ($44.5<{\rm log}~L_{\rm
X}<46.0$) is restricted to type 1 AGN only.  To account for the
optically faint ($r^{\prime}>22.5$), spectroscopically unidentified
X-ray sources, we remove the optical flux limit and implement the
method of \citet{co03} and \citet{ba03} to measure upper limits to the
space density.  We have assigned all unidentified sources to each
feasible $L-z$ bin with the redshift and luminosity set to the mean
value.  With this method, a specific source usually falls within
multiple $L-z$ bins.

At $z<1$, the low luminosity (log~$L_{\rm X}<44.5$) AGN are more than
an order of magnitude more numerous than those with high luminosity,
thus confirming recent results \citep{co03,fi03,ue03}.  This is
evident even when considering the unidentified population since the
upper limit to the space density of luminous AGN (log~$L_{\rm
X}>44.5$) at $z<1$ is fairly constrained.  From Figure~\ref{champ_lz},
we see that most of the sources contributing to the space density at
$z<1$ are non-BLAGN.  These results agree with those reported by
\citet{st03} that show type 2 AGN as the dominant population at
$0.5<z<1$.  The peak values of the two lower luminosity curves
($n\sim3-7\times10^{-5}$ Mpc$^{-3}$) agree with those from the soft
(0.5--2.0 keV) X-ray luminosity function \citep{ha05}.\footnote{Also
reproduced in Figure 8a of a review article on deep extragalactic
X-ray surveys by \citet{br04}.}  We find the space
density a factor of $\sim$3 less, for AGN with $43.0<{\rm log}~L_{\rm
X}<44.5$ at $z\sim0.8$, than the hard (2-10 keV) X-ray survey results
of \citet{ue03}. These comparisons seem reasonable since our
broad band (0.3-8.0 keV) selected sample will be affected by
absorption as known to occur in samples selected in the soft band
(0.5-2.0 keV).

For the luminous AGN (log~$L_{\rm X}>$ 44.5), we measure a rise and
fall of the number density with a peak at $z\sim2.5$.  With 14 AGN in
the high redshift bin ($3<z<4.2$), we have evidence for a drop in the
density that is inconsistent (at $3.4\sigma$) with the peak value
found at $2<z<3$.  This result is not strongly affected by our bright
optical magnitude limit since most AGN at $z>3$ do not all fall near
the flux limit as shown in Figure~\ref{xo:adapt}.  However, given the
large numbers of unidentified soures, the upper limits to the space
density at $z>2$ (Figure~\ref{bxlf:density}) show considerable
uncertainty.  A higher fraction of identified X-ray sources is
required to determine if a decline is present with high significance.
The peak in the space density appears to shift to lower redshift with
decreasing luminosity as modelled by luminosity-dependent density
evolution (\citealt{ue03,mi00}), in contrast to 'pure' luminosity
evolution reported by the optical surveys of bright quasars
(e.g. \citealt{cr04}).

\subsection{Soft band (0.5--2.0 keV): Chandra + ROSAT}
\label{sxlf}

\begin{figure}
\epsscale{1.1}
\plotone{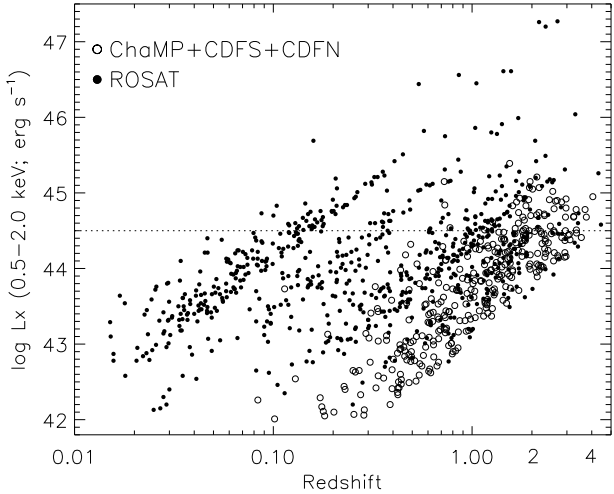}
\caption{X-ray luminosity vs. redshift of 1004 AGNs selected
in the soft (0.5--2.0 keV) band with {\em ROSAT} and {\em Chandra}.  The
horizontal line shows our threshold for measuring the space density of
highly luminous AGN.}
\label{lz_soft}
\end{figure}

With $Chandra's$ broad energy range, we can select an AGN sample in
the soft band (0.5--2.0 keV) to directly compare with the {\em ROSAT}
results.  The ChaMP AGN, supplemented by those in the {\em Chandra} Deep
Fields, can be combined with the {\em ROSAT} sample
\citep{mi00} to measure the number density of AGN in the soft band
with significant numbers of type 1 AGN up to $z\sim4$.  The {\em ROSAT}
sample is a compilation of AGN found in surveys spanning the pencil
beam (0.3 deg$^2$), deep Lockman Hole \citep{le01} to the
wide area ($2.0\times10^4$ deg$^2$), shallow {\em ROSAT} Bright survey
\citep{sc00}.  To directly compare with the published results
from the {\em ROSAT} and optical surveys such as the 2dF \citep{cr04},
SDSS \citep{fa01}, and COMBO-17 \citep{wo03}, we use cosmological
parameters $\Omega_{\rm M}=1$, $\Omega_{\Lambda}=0$, H$_{\circ}=50$ km
s$^{-1}$ Mpc$^{-1}$.  The luminosity is calculated in the observed
frame (no $k$-correction) to compare with the {\em ROSAT} results
\citep{mi00}.  We have assembled a total of 1004 AGN with $L_{\rm
0.5-2.0~keV}>10^{42}$ erg s$^{-1}$, $f_{\rm{X}}>2\times10^{-15}$ erg
cm$^{-2}$ s$^{-1}$, and $r^{\prime}<$ 22.5 (Figure~\ref{lz_soft}).
The ChaMP AGN boost the numbers at high redshift and lower luminosity
due to $Chandra's$ faint limiting flux.  Following the analysis of
\citet{mi00}, we measure the co-moving space density for luminous
(log~$L_{\rm 0.5-2.0~keV}>44.5$) AGN to ensure that we are sensitive to 
these objects out to $z\sim4$.  This minimum luminosity ensures that
our measurements are not highly biased by our X-ray flux limit.

\begin{figure}
\epsscale{1.2}
\plotone{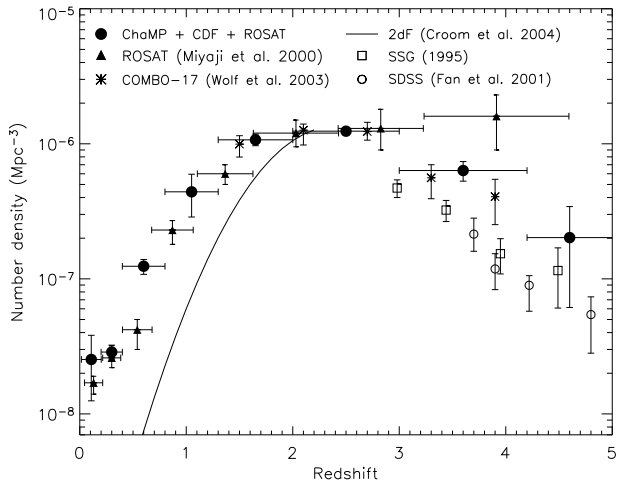}
\caption{Co-moving space density of 217 {\em Chandra} + {\em ROSAT} AGN
selected in the soft (0.5--2.0 keV) band with log~$L_{\rm X}>44.5$
compared to the optical surveys.  The optical space densities have
been scaled to match the X-ray points at $z=2.5$ for ease of
comparison.}
\label{sxlf:fig}
\end{figure}

In Figure~\ref{sxlf:fig}, we plot the co-moving space density of the
217 most X-ray luminous AGN.  With these highly luminous AGN, we can
compare the space density to past {\em ROSAT} results and optical
surveys out to high redshift ($z<5$).  To easily compare the space
density evolution from optical surveys, in Figure~\ref{sxlf:fig}, we
have renormalized their curves to match the ChaMP at $z=2.5$.  For all
but the COMBO-17 survey, optical space densities ($M_B<-26$) have been
scaled up by a factor of 3.2.  COMBO-17 survey, which reaches much
greater depths ($R<24$) was scaled down by a factor of 7.1.  Some
caution must be taken when comparing to optical surveys with vastly
different magnitude limits.  For example, the AGN in the 2dF survey
evolve faster than those from X-ray selected samples at $z<2$
(Figure~\ref{sxlf:fig}).  The COMBO-17 survey, which probes similar
absolute magnitudes (M$_{\rm B}<-22.2$) as the ChaMP, may show
evolution rates comparable to the X-ray surveys over the redshift
interval $1.5<z<2.1$.

The number density of our combined sample is similar to the {\em ROSAT}
results with $z<3$.  Our larger $z>3$ sample (12), while statistically
consistent with the earlier results of \citet{mi00}, has smaller
errors and shows a decrease in the space density from $2<z<3$ to
$3<z<4.2$ significant at the 4.6$\sigma$ level.  Even with the
addition of {\em ROSAT} AGN, we have a slightly smaller sample at $z>3$
than in the broad band sample (Section~\ref{bxlf}) due to lower
fluxes, luminosity or counts in the soft band for the ChaMP AGN.  We
find reasonable agreement with a slight excess between the relative
drop in the space density of the X-ray and optical surveys at $z>2$;
our data are consistent with a decline in the X-ray luminosity function
similar to that observed in the optical.

\section{Conclusion}

We have measured the broad band (0.3--8.0 keV) co-moving space density
using a sample of 368 X-ray emitting AGN detected by {\em Chandra}.
Our spectroscopic magnitude limit to date allows 
the inclusion of optically obscured AGN to $z\leq 1$ and unobscured AGN out to
$z\sim4$.

Our primary results are as follows:

\begin{itemize}

\item We confirm that low luminosity AGN (log~$L_{\rm X}<$ 44.5)
are more prevalent at $z<1$ than higher luminosity AGN, as seen by
\citet{co03}, \citet{fi03}, and \citet{ue03}.
Non-BLAGN are the major contributor to the co-moving space density at
these low redshifts, as reported by \citet{st03}.

\item The space density of type 1, highly luminous (log~$L_{\rm X}>44.5$)
X-ray selected AGN rises from the present epoch to a peak at
$z\sim2.5$ and then declines at $z>3$.  This behavior is similar to
that of optically selected surveys.  This is the first X-ray selected
survey to detect a decline at high redshifts.  The evolution is
evident using a sample of 311 AGN from the ChaMP selected in the broad
(0.3--8.0 keV) {\em Chandra} band and 217 luminous QSOs selected in
the soft band (0.5--2.0 keV) from {\em Chandra} and {\em ROSAT}.

\end{itemize}

Our results support a more rapid depletion of fuel for the high
luminosity AGN from $z\sim2.5$ to the present epoch \citep{me04},
perhaps due to their higher accretion rate.  At $z<1$, the lower
luminosity AGN evolve slowly, which may be attributable to a low
accretion rate, compared to the more luminous AGN.  At $z>3$, the
decline in the space density of highly luminous AGN represents the
growth phase of supermassive black holes during a period of rapid
galaxy assembly.  The uncertainties in the X-ray luminosity function
will improve with larger samples and a higher fraction of source
classification, either through spectroscopic or photometric techiques.
With the inclusion of obscured, highly luminous QSOs, X-ray surveys
are well on the way to presenting a more comprehensive view of AGN
evolution.

\acknowledgments

We are greatly indebted to NOAO and the SAO TACs for their support of
this work.  We thank the staffs at KPNO, CTIO, Las Campanas,
W. M. Keck Observatory, FLWO, and MMT for assistance with optical
observations.  We thank John Huchra for his suggestions on dealing
with complex selection effects.  We also thank Takamitsu Miyaji and
Guenther Hasinger for providing us with the {\em ROSAT} AGN catalog.

We gratefully acknowledge support for this project under NASA CXC
archival research grants AR3-4018X and AR4-5017X.  TLA, RAC, PJG, DWK,
AEM, HT, and BW also acknowledge support through NASA Contract NASA
contract NAS8-39073 (CXC).

\end{document}